\documentclass[twocolumn,amsmath,amssymb,showpacs,prb]{revtex4-1}

\usepackage[pdftex]{graphicx} 
\usepackage{epstopdf}
\usepackage{verbatim}

\newcommand{\ket}[1]{\ensuremath{|#1\rangle}}
\newcommand{\bracket}[2]{\ensuremath{\langle#1|#2\rangle}}

\newcommand{\avg}[1]{\ensuremath{\langle #1 \rangle}}

\newcommand{\defequal}{\stackrel{\text{\scriptsize def}}{=}}

\def\be{\begin{equation}}
\def\ee{\end{equation}}
\def\bea{\begin{eqnarray}}
\def\eea{\end{eqnarray}}

\usepackage{color}

\begin{document}

\title{Real-Space Parallel Density Matrix Renormalization Group}

\author{E.\,M.\ Stoudenmire}
\affiliation{Department of Physics and Astronomy, University of California, Irvine, CA 92697}
\author{Steven R.\ White}
\affiliation{Department of Physics and Astronomy, University of California, Irvine, CA 92697}

\date{\today}

\begin{abstract}
We demonstrate how to parallelize the density matrix renormalization group (DMRG) algorithm
in real space through a straightforward modification of serial DMRG.
This makes it possible to apply at least an order of magnitude more computational power to challenging
simulations, greatly accelerating investigations of two-dimensional systems and large parameter spaces.
We discuss details of the algorithm and present benchmark results including a study of
valence-bond-solid order within the square-lattice Q$_2$ model and N\'eel order within the triangular lattice 
Heisenberg model.
The parallel DMRG algorithm also motivates an alternative canonical form for matrix product states.
\end{abstract}

\pacs{
05.10.Cc, 
02.70.-c, 
05.30.-d 
}

\maketitle

\section{Introduction}

The density matrix renormalization group (DMRG) is a 
method for computing ground states of one-dimensional (1d) systems \cite{White:1992,*White:1993a, Schollwoeck:2005, Noack:2005} 
which has recently proven surprisingly effective for studying two-dimensional (2d) models, 
especially those beyond the reach of quantum Monte Carlo due to frustratated interactions or mobile fermions.\cite{White:2007,Jiang:2011,Yan:2011,Corboz:2011,Jiang:2012s,Bauer:2012,Depenbrock:2012,Cincio:2013,Zhu:2013,Ganesh:2013,Stoudenmire:2012a}
DMRG is also valuable in quantum chemistry where it has extended the ability of
previous methods to deal with strong correlation.\cite{Kurashige:2011,Chan:2011,Sharma:2012}
But applying DMRG to 2d and quantum chemical systems is computationally very demanding.
Even in 1d, DMRG can be costly for systems having many degrees of freedom.\cite{Hachmann:2006,Dolfi:2012,Stoudenmire:2012d}
Finding an efficient way to divide a single DMRG calculation across multiple processors 
would make larger calculations tractable, but a truly
parallel DMRG algorithm has remained an outstanding problem.

One previous approach to parallelizing DMRG has been to 
parallelize over different terms in the Hamiltonian.\cite{Chan:2004} This is effective in a 
quantum chemistry context where the number of terms is especially large. 
However, since the terms contained entirely within a single
spatial region are already combined as much as possible within DMRG, 
the efficiency of this approach is less than ideal.
A similar type of limited parallelism subdivides matrices into quantum
number blocks.\cite{Hager:2004,Kurashige:2009} But this approach is restricted by a limited range of quantum
numbers and a large variation in subblock size. The least powerful, but simplest form
of parallelism breaks up dense matrix computations into subblocks and is performed
automatically by standard linear algebra libraries.

\begin{figure}[tp]
\includegraphics[width=\columnwidth]{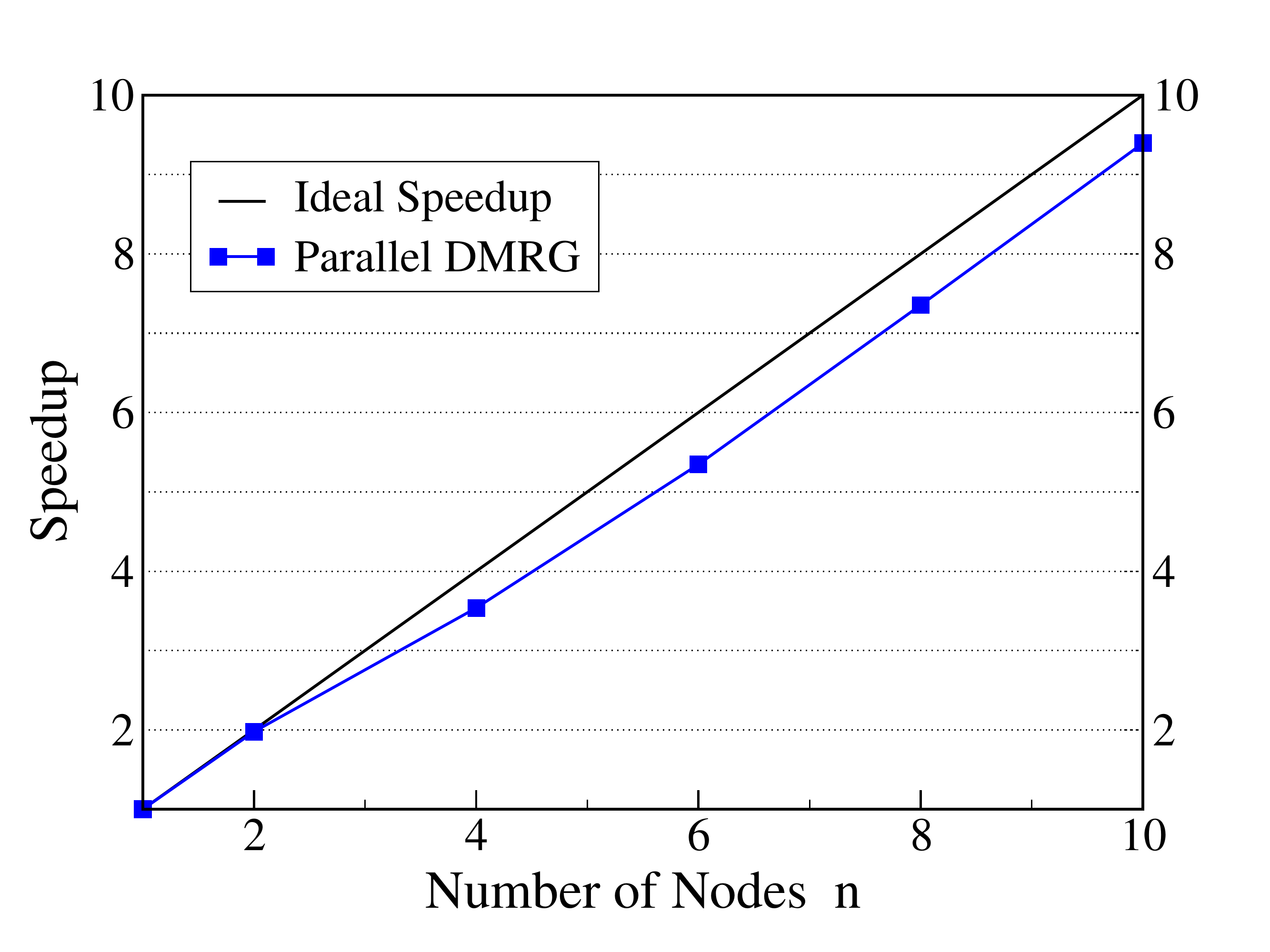}
\caption{Timing of parallel DMRG ground state calculations for the spin $1/2$ Heisenberg model on the $24\times 8$ square lattice 
(cylindrical boundary conditions \cite{Stoudenmire:2012a}).
A speedup of $S$ indicates that the calculation using $n$ nodes was $S$ times faster than the same calculation using only one node.
Each calculation consisted of 10 sweeps and reached a relative energy accuracy of $10^{-5}$ by keeping $m=2000$ states in the final two sweeps.
The speedup is so close to ideal for the two-node case because the reflection symmetry permits optimal load balancing and minimizes 
waiting time at the communication step.}
\label{fig:speedup}
\end{figure}

Here we present a much more powerful form of parallelism, dividing a single DMRG calculation
over separate regions of the system in real space.
Except at the boundary of each region, 
the algorithm reduces to standard finite-size DMRG,
making it relatively straightfoward to implement in existing, highly-optimized DMRG codes.
Real-space parallelism can be used independently
of the other types discussed above and becomes increasingly effective as the system size increases.
In practice, we observe close to ideal speedups as shown in Fig.~\ref{fig:speedup}.

\begin{figure*}[t]
\includegraphics[width=450px]{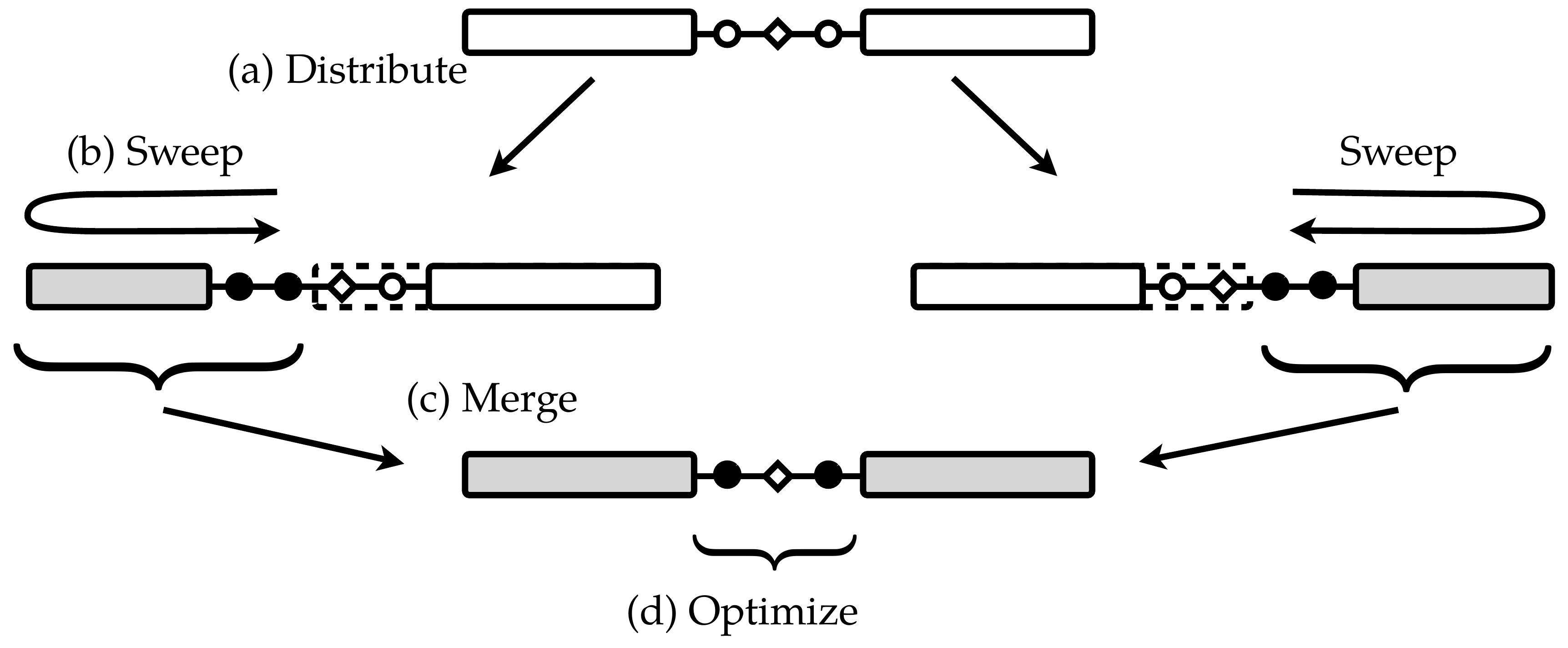}
\caption{
The parallel DMRG algorithm for the simplest case of dividing a single calculation over two machines.
Circles represent lattice sites (or matrix product state site-tensors) and the diamonds each represent the same matrix $V$ on the shared bond as in Eq.~(\ref{eqn:isvd}).
In step (a), the local information needed to sweep left is copied to the left machine and similarly for the right.
Each machine then performs a DMRG sweep (b) in parallel over its half of the system.
In step (c) the wavefunctions are merged together using the prescription preceding Eq.~(\ref{eqn:merge}).
Before repeating the algorithm, the merged state is optimized (d) using Lanczos or Davidson on the shared bond.
}
\label{fig:algorithm}
\end{figure*}

Note that some extensions of DMRG are already real-space parallelizable, such as time-dependent DMRG algorithms based
on factorizations of the time-evolution operator, as in a Suzuki-Trotter approximation.\cite{Daley:2004, White:2004r} 
These algorithms can be used to compute ground states
through imaginary time evolution. But imaginary time evolution is much less efficient than the usual
DMRG diagonalization-based method, therefore it is primarily useful when the standard DMRG ground state
approach is not applicable (for example, when optimizing 2d tensor networks such as PEPS \cite{Xie:2009,Pizorn:2011,Czarnik:2012}).

After introducing the parallel DMRG algorithm in section \ref{section:algorithm}, we use it to
compute the strength of the valence-bond-solid order in the pure $Q_2$ model on the square lattice (section \ref{section:pureQ2}) and the strength of the Ne\'el order
in the Heisenberg model on the triangular lattice (section \ref{section:triangular}). Both applications involve calculations which would take many
weeks using serial DMRG but require only a few days using our parallel approach. Finally, in section \ref{section:icmps} we discuss how real-space parallel DMRG motivates an alternative canonical gauge for matrix product states.

\section{Parallel DMRG Algorithm \label{section:algorithm}}

To describe the algorithm, first consider parallelizing a single DMRG calculation over just two regions. 
For concreteness, take the system to have six sites such that the center bond connects sites 3 and 4.
As a warmup, first perform a few sweeps of the standard, serial finite-size 
DMRG algorithm \cite{White:1992,*White:1993a, Schollwoeck:2005, Noack:2005} keeping only a small number
of states so this non-parallel part takes relatively little time. Stop when the two exposed sites are at the center bond as in Fig.~\ref{fig:algorithm}(a).


\begin{figure}[b]
\includegraphics[width=\columnwidth]{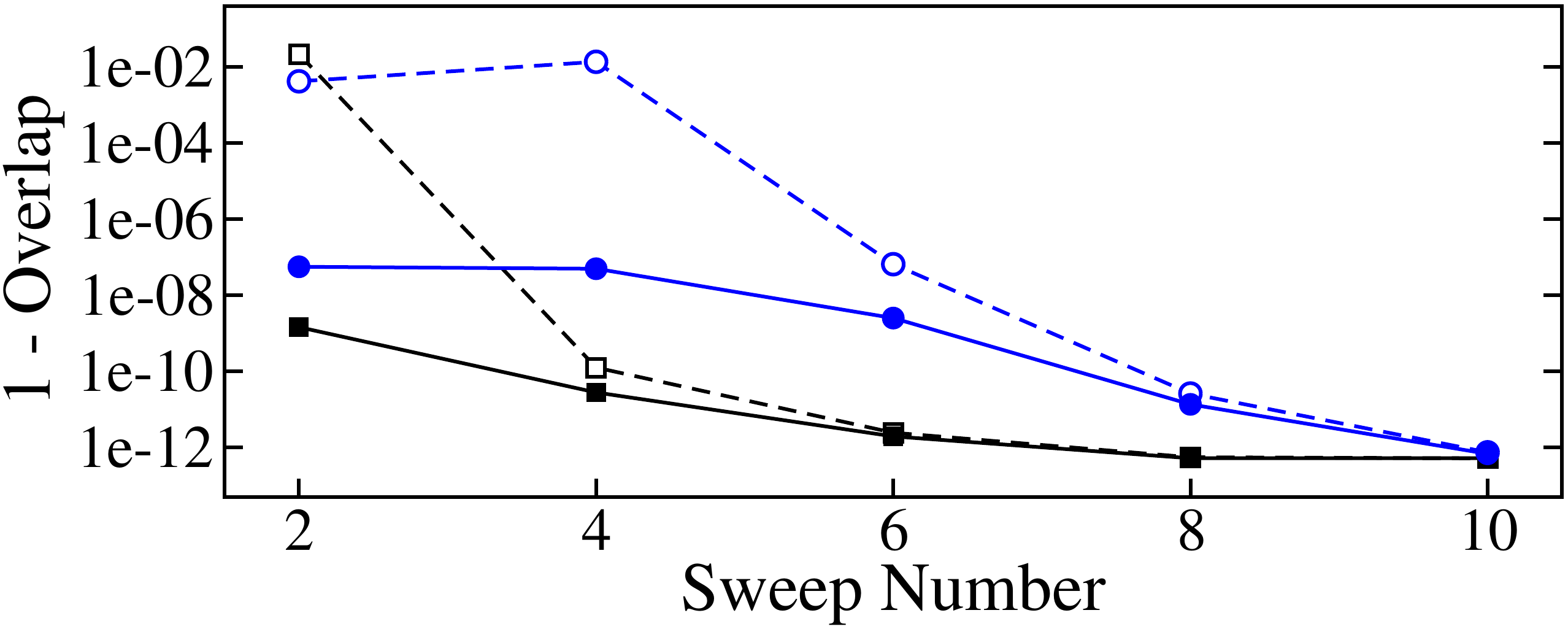}
\caption{Prediction fidelity $1-\bracket{\Psi^\prime_n}{\Psi_n}$ at the center (shared) bond of a DMRG calculation parallelized across two nodes.
\ket{\Psi^\prime_n} is the initial merged wavefunction following sweep $n$ as in Eq.~(\ref{eqn:merge}). \ket{\Psi_n} is 
the wavefunction resulting from a fully converging Davidson calculation initialized with $\ket{\Psi^\prime_n}$.
The systems are, from top to bottom, the 200 site $S=1/2$ Heisenberg chain without warmup sweeps (dashed blue circles); 
 $S=1/2$ chain with warmup sweeps (solid blue circles); $S=1$ Heisenberg chain without (dashed black squares) 
and with warmup sweeps (solid black squares).
The warmup consisted of 5 serial DMRG sweeps keeping $m=50$ states. For the parallel sweeps, $m$ was increased after every other sweep
to a maximum of $600$.
}
\label{fig:overlap}
\end{figure}

At this point the wavefunction within DMRG has the form
\be
\ket{\Psi} = \sum_{\alpha_2 s_3 s_4 \alpha_4} \Psi^{\alpha_2 s_3 s_4 \alpha_4} \ket{\alpha_2}_L \ket{s_3} \ket{s_4} \ket{\alpha_4}_R
\label{eqn:wf}
\ee 
where $s_3, s_4 = 1,\ldots,d$ label the lattice basis on sites 3 and 4, and 
$\alpha_2, \alpha_4 = 1,\ldots,m$ label orthonormal many-body states approximating
$\ket{\Psi}$ within the left and right blocks, respectively.

DMRG proceeds in two steps: first, this wavefunction is optimized using a few Lanczos or Davidson steps
with the Hamiltonian projected into the $\ket{\alpha_2}_L \ket{s_3} \ket{s_4} \ket{\alpha_4}_R$ basis.\cite{Noack:2005} 
Then a renormalization group procedure is carried out based on
the singular value decomposition (SVD) of the amplitudes $\Psi$ 
\be
\Psi^{(\alpha_2 s_3) (s_4 \alpha_4)} = \sum_{\alpha_3} A^{(\alpha_2 s_3)}_{\alpha_3} \Lambda^{\alpha_3} B^{(s_4 \alpha_4)}_{\alpha_3} \:.
\label{eqn:svd}
\ee
(with $\Psi$ temporarily treated as an $(md)\times(dm)$ matrix).
Following the SVD, all but the largest $m$ singular values are truncated.

In the serial DMRG algorithm, one next selects either the transformation matrix $B$ to grow the right block, in which case the product 
$A\Lambda$ is used to compute the wavefunction amplitudes on the bond to the left, or one chooses the transformation matrix
$A$ to grow the left block and $\Lambda B$ to form the next wavefunction amplitudes.\cite{White:1996,Schollwoeck:2011}
Iterating these steps generates a sweeping procedure which improves the wavefunction at every bond sequentially.

Now assume we have two computers able to work in parallel. Following the SVD, the first machine could sweep left 
while the second simultaneously sweeps right.
However this soon leads to a conceptual issue: each computer will be working with a different wavefunction globally.
One can temporarily ignore this problem, but when the computers meet again it will be unclear how
to merge their wavefunctions for the purpose of performing a DMRG step on their shared bond.

To overcome this inconsistency, rewrite Eq.~(\ref{eqn:svd}) but insert the identity $\Lambda V = 1$, where $V\defequal\Lambda^{-1}$:
\begin{align}
\Psi^{\alpha_2 s_3 s_4 \alpha_4 } & = \sum_{\alpha_3} A^{\alpha_2 s_3}_{\alpha_3} \Lambda^{\alpha_3} V_{\alpha_3} \Lambda^{\alpha_3} B^{s_4 \alpha_4}_{\alpha_3} \label{eqn:inverse} \\
& \defequal \sum_{\alpha_3} \ \psi_3^{\alpha_2 s_3 \alpha_3}\  V_{\alpha_3}\  \psi_4^{\alpha_3 s_4 \alpha_4}  \label{eqn:isvd} \:.
\end{align}
Again both machines can sweep in parallel, but when they return to their shared bond---the left with updated amplitudes $\psi^\prime_3$ and the right with amplitudes $\psi^\prime_4$---there is a consistent way to merge the two wavefunctions. 
To define the merged wavefunction, take the DMRG basis states $\ket{\alpha_2}_L$ for the left from the machine sweeping over that region;
the states $\ket{\alpha_4}_R$ for the right from the other machine; and for the amplitudes at the center bond choose
\be
\Psi^\prime = \psi^\prime_3\ V_3\ \psi^\prime_4 \:,
\label{eqn:merge}
\ee
similarly to Eq.~(\ref{eqn:isvd}), using the original  $V$.
Note that the exact ground state is a fixed point of this procedure.
After merging the two wavefunctions for optimization on the shared bond, the merged wavefunction can again be split in two if more parallel sweeps are needed.

Though there is no communication between machines prior to each merge, in practice we find that
Eq.~(\ref{eqn:merge}) provides a good initial state for the Lanczos or Davidson steps on the shared bond, as shown in Fig.~\ref{fig:overlap}.
However, because each machine otherwise updates the wavefunction independently, DMRG convergence is typically slower at the shared bond, especially 
near the beginning of a calculation.
Although this means parallel DMRG gives slightly worse results compared to serial DMRG for the same number of sweeps,
the nearly ideal speedup in calculation time more than compensates for this effect.

Though the discussion above emphasizes the wavefunction, an important part of DMRG is transforming any projected operators (such as
the Hamiltonian)
while sweeping. First, as each machine sweeps away from the shared bond in parallel, this transformation occurs in the usual way.
For example, the matrix $B$ from Eq.~(\ref{eqn:svd}) transforms the Hamiltonian into the local basis of the next pair of sites to the left.
Later when the two machines merge their wavefunctions, they merge operators in an analogous way: operator terms acting in the left
half of the system are approximated by their projection into the basis states from the left machine and similarly for the right.

The role of the matrix $V$ in the merge can be understood by observing that DMRG approximately preserves the reduced density matrix 
over regions where it does not sweep. 
Assuming that the wavefunction does not change too drastically after each sweep,
the matrix $V$ normalizes $\psi^\prime_4$ on the right such that it approximates 
the reduced density matrix eigenstates on the right half of the system, and similarly for the left.
Note the resemblance of Eq.~(\ref{eqn:merge}) to the prediction step in the infinite DMRG algorithm of Ref.~\onlinecite{McCulloch:2008}, which directly
motivated the present work.
As explained in Section~\ref{section:icmps}, from a matrix product state point of view Eq.~(\ref{eqn:inverse}) is a transformation from a gauge having only 
one orthogonality center (a tensor whose indices all label orthonormal states) to a gauge with two orthogonality centers.

One concern about introducing a matrix $V=\Lambda^{-1}$ at shared bonds is that the relative errors made in computing small singular values
may be amplified when inverting them to compute the elements of $V$, which can be orders of magnitude larger than one.
To address this issue, we have implemented an accurate SVD algorithm described in detail in the Appendix. 
By recursively performing an SVD on submatrices containing only the smallest singular values, one can 
obtain uniform relative accuracy for all singular values---not just the largest.

\begin{figure}[t]
\includegraphics[width=\columnwidth]{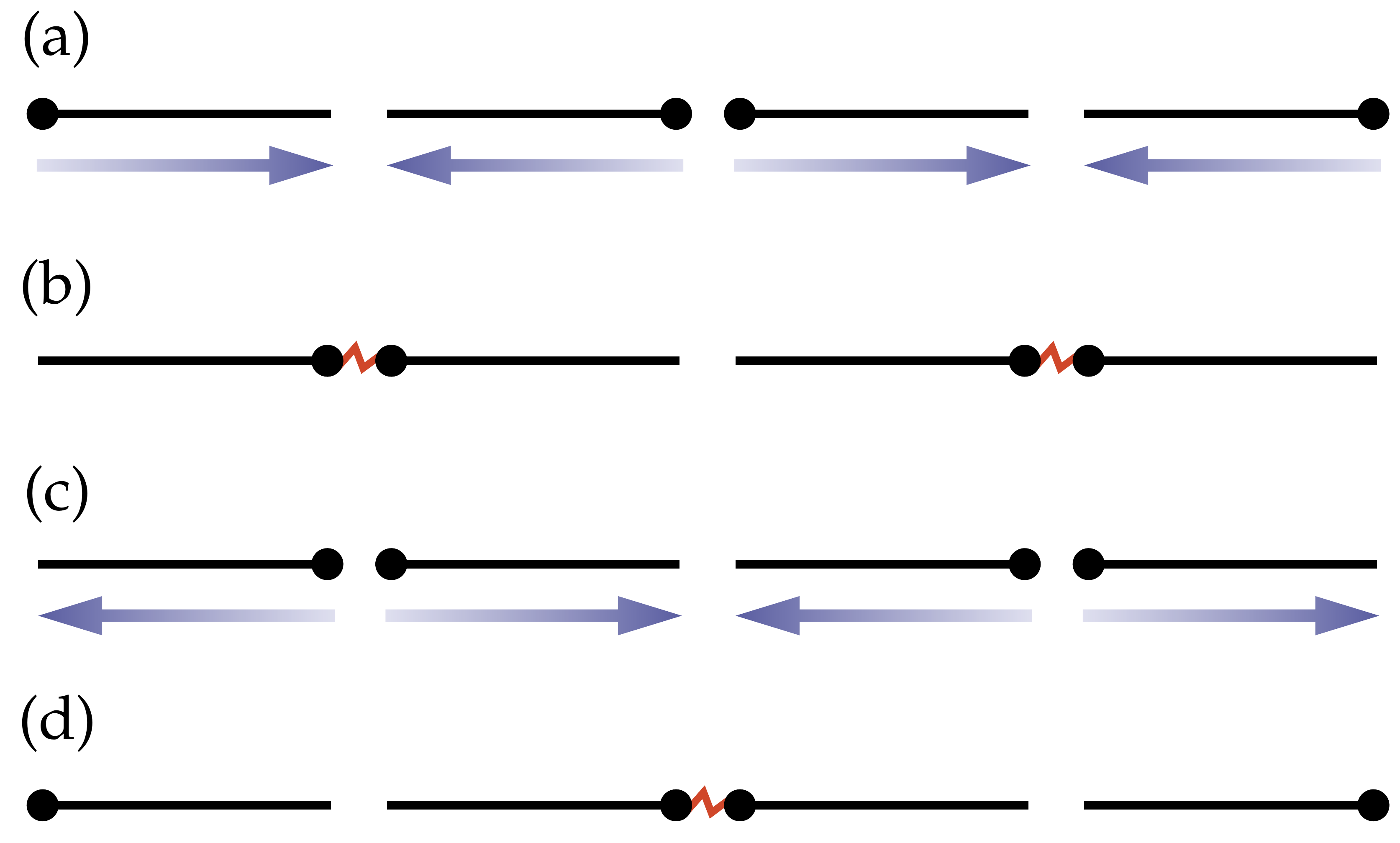}
\caption{Sweeping pattern for one full sweep of the parallel DMRG algorithm split over four computational nodes. First, (a) the nodes are positioned
in a spatially staggered pattern and sweep to the other end of their block. When the nodes reach the end of their block (b) they wait for their
neighboring node to arrive then communicate. Finally the nodes sweep back (c) to their starting positions and (d) communicate with their other neighbor.
}
\label{fig:sweeping_pattern}
\end{figure}

The two-block algorithm described above can be readily extended to $n$ real-space blocks by first splitting the system in two, 
then further splitting each sub-block until there are $n$ total. This motivates the sweeping pattern shown in Fig.~\ref{fig:sweeping_pattern}.
Odd numbered nodes start on the left end of their block and even nodes on the right. This way, when a node reaches the end
of its block the next node is ready to communicate. If a node reaches the end of its block before its
neighbor arrives, it is better for the node to wait instead of immediately beginning the next half sweep.
Having an updated environment far outweighs the loss in efficiency due to a node briefly remaining idle.

\section{Benchmark Applications \label{section:applications}}

\subsection{Pure Q$_2$ Model \label{section:pureQ2}}

To demonstrate that real-space parallel DMRG can be used to accelerate very challenging
two-dimensional DMRG calculations, we use it to study
the $S=1/2$, pure Q$_2$ model on the square lattice. 
This model has been proposed as a benchmark for testing the predictive ability of DMRG for 2d systems.\cite{Sandvik:2012}
Extensive quantum Monte Carlo (QMC) calculations show the model is in a phase with weak columnar valence bond solid (VBS) order.\cite{Sandvik:2007,Sandvik:2010d,Sandvik:2012} Because DMRG is limited to much smaller finite-size systems than QMC, a reasonable concern is that DMRG could miss such weak order 
and possibly mistake the phase for a spin liquid.

\begin{figure}[t]
\includegraphics[width=\columnwidth]{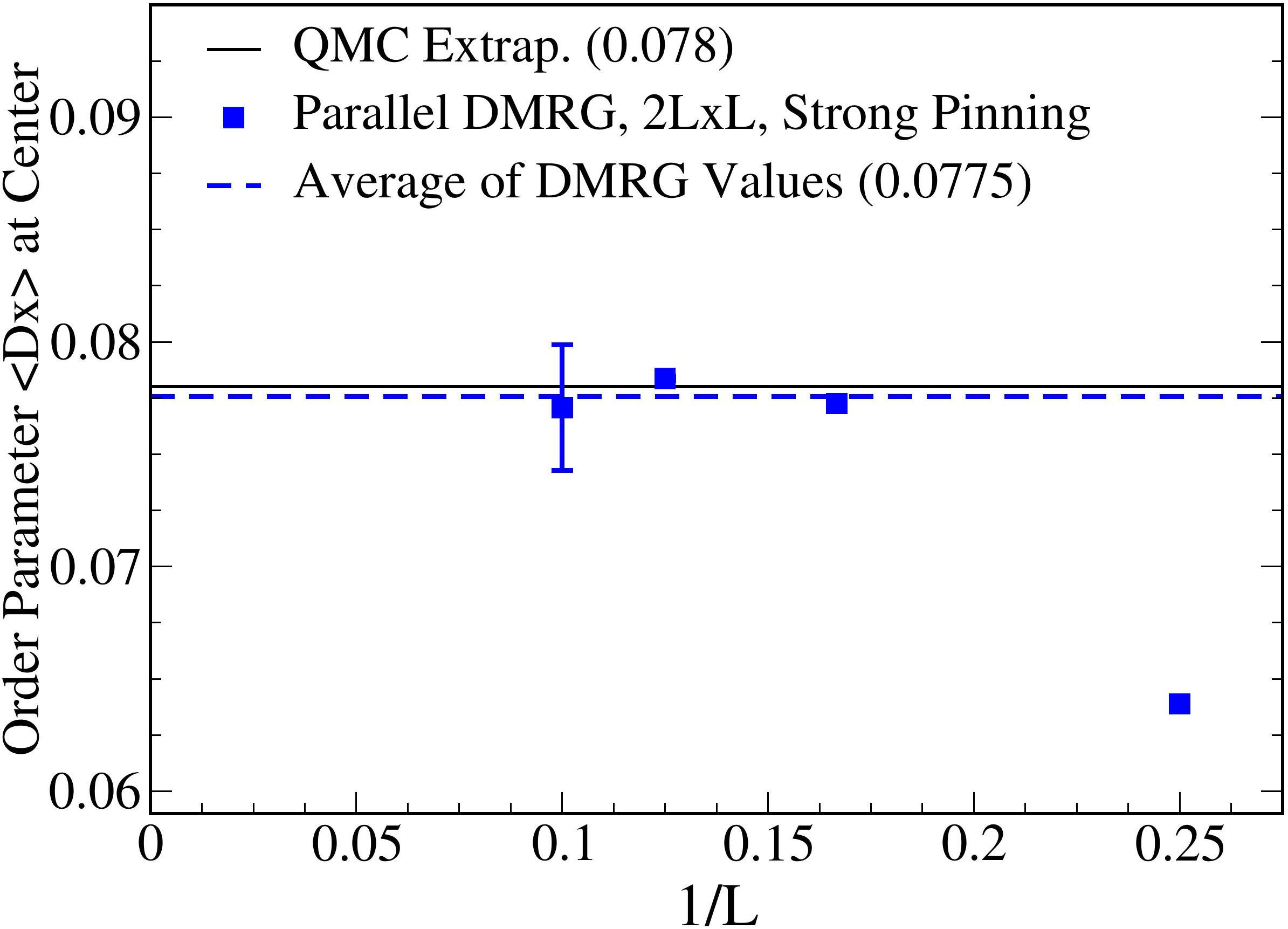}
\caption{Columnar VBS order parameter $\avg{\text{D}_x}$ on open cylinders of the square-lattice $Q_2$ model Eq.~(\ref{eqn:q2}) with
strong-pinning boundary conditions.\cite{Stoudenmire:2012a} Each order parameter value was estimated by increasing the states kept $m$ after every other sweep 
and extrapolating in the truncation error. Error bars are smaller than symbol sizes except for the $L=10$ system.}
\label{fig:Dx}
\end{figure}

The Hamiltonian of the pure Q$_2$ model is defined as
\begin{equation}
H = -Q_2 \sum_{\langle i j ; k l \rangle} (\mathbf{S}_i\cdot\mathbf{S}_j - 1/4) (\mathbf{S}_k\cdot\mathbf{S}_l - 1/4) \label{eqn:q2}
\end{equation}
where the sum is over all pairs of bonds $(i,j)$ and $(k,l)$ on opposite sides of the elementary square plaquettes.
To investigate the columnar VBS order following Ref. \onlinecite{Sandvik:2012}, 
we study finite-size systems of width $L$ in the $y$ direction and length $2L$ in the $x$ direction.
For each system size, we measure the order parameter
\be
\avg{\hat D_x} = \frac{1}{L} \sum_{y} \avg{\mathbf{S}_{\mathbf{r}}\cdot\mathbf{S}_{\mathbf{r}+\hat x}} - \frac{1}{2} \avg{\mathbf{S}_{\mathbf{r}-\hat x}\cdot\mathbf{S}_\mathbf{r}} - \frac{1}{2} \avg{\mathbf{S}_{\mathbf{r}+\hat x}\cdot\mathbf{S}_{\mathbf{r}+2\hat x}}
\ee
averaged over all $\mathbf{r}=(L,y)$ at the central column of the system.

In the shorter $y$ direction we take periodic boundary conditions but use open boundary conditions in the $x$ direction
for technical reasons. In contrast to Ref. \onlinecite{Sandvik:2012}, however, we
include extra Hamiltonian terms at the edges to pin columnar VBS order.
Following the strong-pinning prescription of Ref.~\onlinecite{Stoudenmire:2012a}, 
we imagine fictitious spins just beyond the edge of the system
locked into an ideal columnar VBS in the $x$ direction. Tracing over these fictitious spins induces 
a term
\be
H_{\text{pin},\mathbf{r}} = \frac{Q_2}{4} (\mathbf{S}_{\mathbf{r}}\cdot\mathbf{S}_{\mathbf{r}+\hat y} - 1/4)
\ee
on each vertical bond along the edges of the real system.
The combination of these terms and a $2\!:\!1$ aspect ratio helps to control finite-size effects.

We carried out parallel DMRG calculations for cylinders of width $L=4,6,8$ and $10$ and show the resulting $\avg{\hat D_x}$ 
order parameter values in Fig.~\ref{fig:Dx}. Each calculation was parallelized over four real-space blocks 
with each block assigned to a separate 8-core Intel Harpertown 2.66~GHz node.
The largest calculation---keeping up to $m=3000$ states for the $L=10$ system---took 6 days and  
would have taken 18--21 days without parallelization. 
All calculations could easily have been parallelized further with access to more nodes.

With our choice of aspect ratio and boundary conditions, we found only a weak dependence of the order parameter on 
system size for large enough $L$. By averaging $\avg{\hat D_x}$ for the largest three systems, 
we estimate a value $\avg{\hat D_x}=0.078(3)$ for the thermodynamic limit, in good
agreement \footnote{Our calculations only agree up to a factor of precisely two. Up to this factor, 
we have successfully reproduced the results (not shown) of Ref.~\onlinecite{Sandvik:2012} when omitting the edge pinning terms. 
We therefore report twice the value stated in Ref.~\onlinecite{Sandvik:2012} (p.~10) as the QMC estimate of the 2d order parameter.} 
with the value $0.078$  predicted by quantum Monte Carlo.\cite[p.\,10]{Sandvik:2012}

\subsection{Triangular Heisenberg Antiferromagnet \label{section:triangular}}

As a second application of parallel DMRG, we measure the $120^\circ$ N\'eel 
order of the antiferromagnetic $S=1/2$ Heisenberg model $H=J \sum_{\avg{i j}} \mathbf{S}_i \cdot \mathbf{S}_j$ 
on the triangular lattice.
In contrast to the Q$_2$ model, this system is beyond the reach of QMC due to the sign problem.
It is challenging even for DMRG because of the high coordination number of the lattice.

Reference~\onlinecite{White:2007} computed the staggered magnetization of this system using DMRG on
cylinders up to width $L_y = 9$, but found somewhat unsatisfactory results for the largest systems. 
In particular, it was unclear whether the order parameter develops a crossing point at some
particular aspect ratio.

\begin{figure}[t]
\includegraphics[width=\columnwidth]{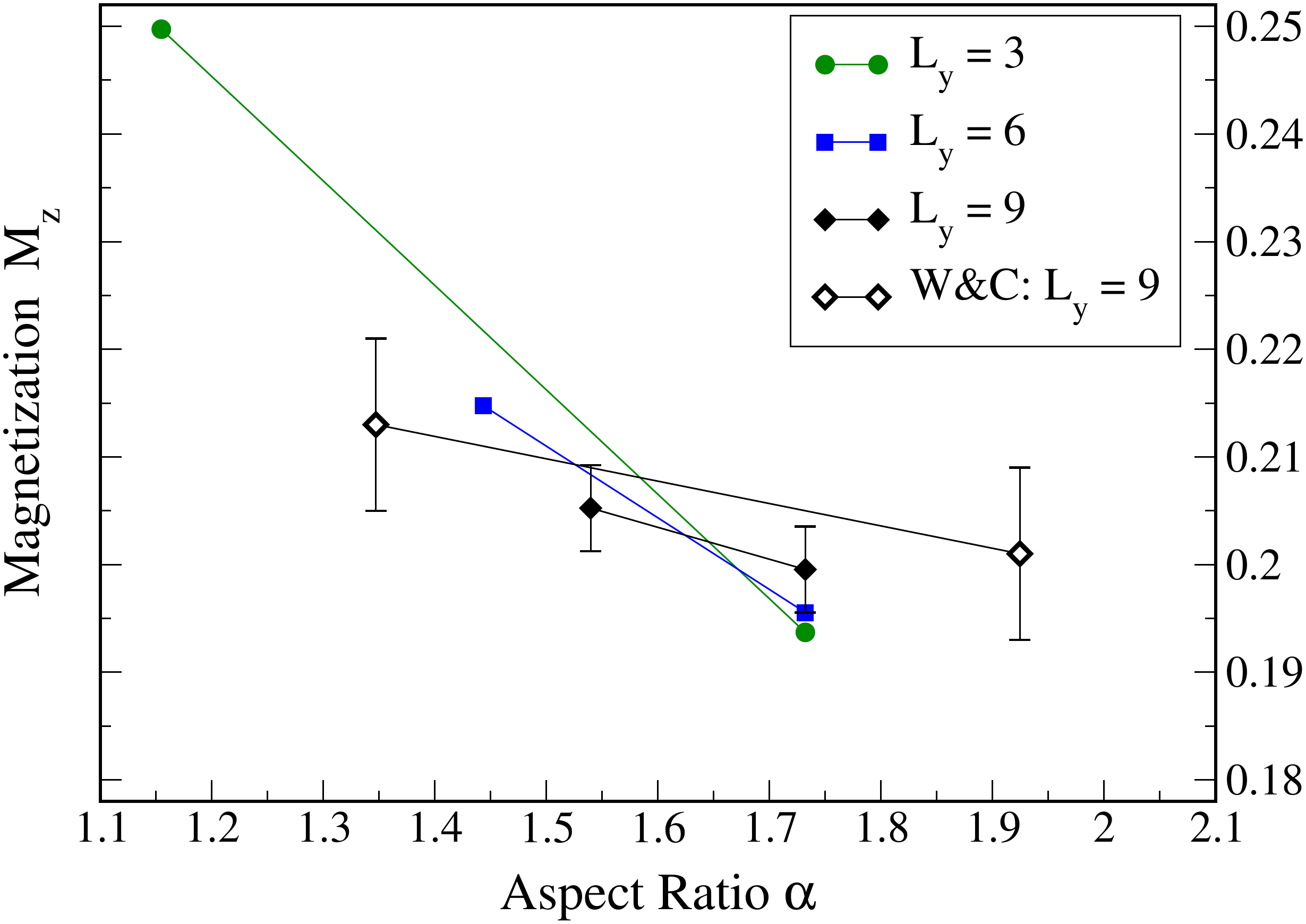}
\caption{Staggered magnetization M$_z$ of the antiferromagnetic $S=1/2$ Heisenberg model on open cylinders of the triangular lattice with
strong-pinning boundary conditions.\cite{White:2007} The aspect ratio of each cylinder is defined as $\alpha=L_x/L_y$. Each magnetization value was computed by gradually increasing the number of states kept
and extrapolating in the truncation error. Error bars are smaller than symbol sizes except for the $L_y=9$ systems.}
\label{fig:tri_mag}
\end{figure}

Here we revisit that calculation using parallel DMRG to study the width $L_y=9$ systems more accurately.
Our procedure,
described in more detail in Refs.~\onlinecite{Stoudenmire:2012a,White:2007}, is to study cylinders with $L_y$ sites and 
periodic boundary conditions in the $y$ direction and $L_x$ sites and open boundaries in the $x$ direction. 
We apply a magnetic field term $-(J/2) \,S^z_i$ to sites $i$ at the open edges on only one of the three sublattices 
in order to pin the magnetization direction and reduce the bulk entanglement.
We then measure $M_z = \avg{S^z}$ on the pinned sublattice averaged over the center column of the cylinder.
First, we reproduce the results of Ref.~\onlinecite{White:2007} for cylinders of width $L_y=3$ and $L_y=6$.
These results are shown in Fig.~\ref{fig:tri_mag} and agree with those of Ref.~\onlinecite{White:2007} essentially exactly.

For the width 9 systems we chose aspect ratios closer to the crossing point than Ref.~\onlinecite{White:2007}.
We kept up to $m=5000$ states in DMRG and extrapolated $M_z$ in the truncation error. 
We also performed a second set of runs increasing the number of states according to a different pattern to test our extrapolation
and found excellent agreement. The largest runs took about a week each using four nodes, and therefore would have taken 
3-4 weeks without paralellization.

As shown in Fig.~\ref{fig:tri_mag}, by keeping more states and choosing aspect ratios closer to the crossing point, we
obtain tighter bounds on the 2d magnetization. By averaging the 
two $L_y=9$ values in Fig.~\ref{fig:tri_mag}, we find $\alpha_c \simeq 1.64$ and estimate $M_z = 0.202(2)$ for the infinite 2d system.

\section{Inverse Canonical Matrix Product State Gauge \label{section:icmps}}

An interesting byproduct of the parallel DMRG wavefunction transformation Eqs.~(\ref{eqn:isvd}) and (\ref{eqn:merge}) is that it motivates 
an alternative matrix product state (MPS) gauge similar to the canonical gauge \cite{Vidal:2003, Schollwoeck:2011} but with every site an
 orthogonality center rather than every bond. By orthogonality center (OC) we mean any tensor whose indices all label orthonormal states.

\begin{figure}[b]
\includegraphics[width=\columnwidth]{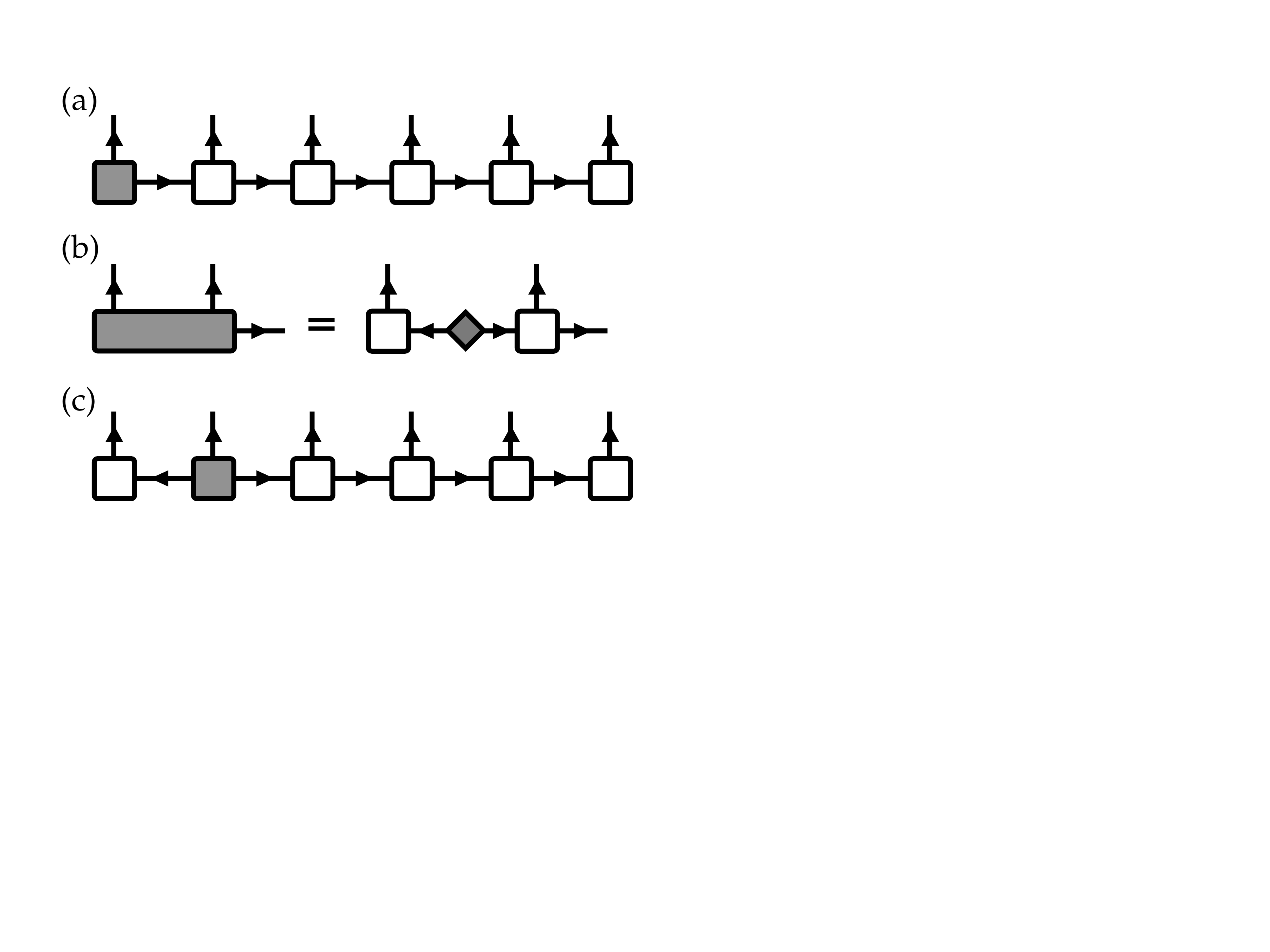}
\caption{Matrix product state in (a) the left-canonical gauge where the OC is the first site.
Combining (b) the first two site tensors, computing a singular-value decomposition, then multiplying
the singular value matrix (shaded diamond) into the second site tensor
transforms the MPS into (c) a mixed-canonical gauge with the second site as the OC.\cite{Schollwoeck:2011}}
\label{fig:mixedcmps}
\end{figure}

For example, the only OC of an MPS in the left-canonical gauge---the gauge naturally occurring at
the end of a full DMRG sweep---is the first site tensor.\cite{Schollwoeck:2011}
This MPS gauge is shown in Fig.~\ref{fig:mixedcmps}(a) with the first site tensor shaded to indicate
it is an OC. The arrow on each index indicates whether that index transforms as a ket (vector) or bra (covector).
An outgoing arrow indicates a ket index and an incoming arrow a bra index.
In traditional tensor notation this corresponds to a raised or lowered index, respectively.
Because the physical site indices transform as kets by definition, they always point out of a ket MPS.

In our convention, the arrows of the virtual or link indices within an MPS do not flow according to a rigid pattern, such as left to right, 
but rather out of the OC (or out of each OC if there are more than one).
To motivate this convention, perform a gauge transformation of the MPS \ref{fig:mixedcmps}(a). Contract the first two site tensors over their shared link index, then
compute an SVD as shown in Fig.~\ref{fig:mixedcmps}(b), and multiply the diagonal singular-value matrix into the second site tensor. The resulting MPS Fig.~\ref{fig:mixedcmps}(c) has its OC at the second site since the first site tensor is formed from a unitary matrix. As expected, the arrows now flow out of the second site tensor.

\begin{figure}[t]
\includegraphics[width=\columnwidth]{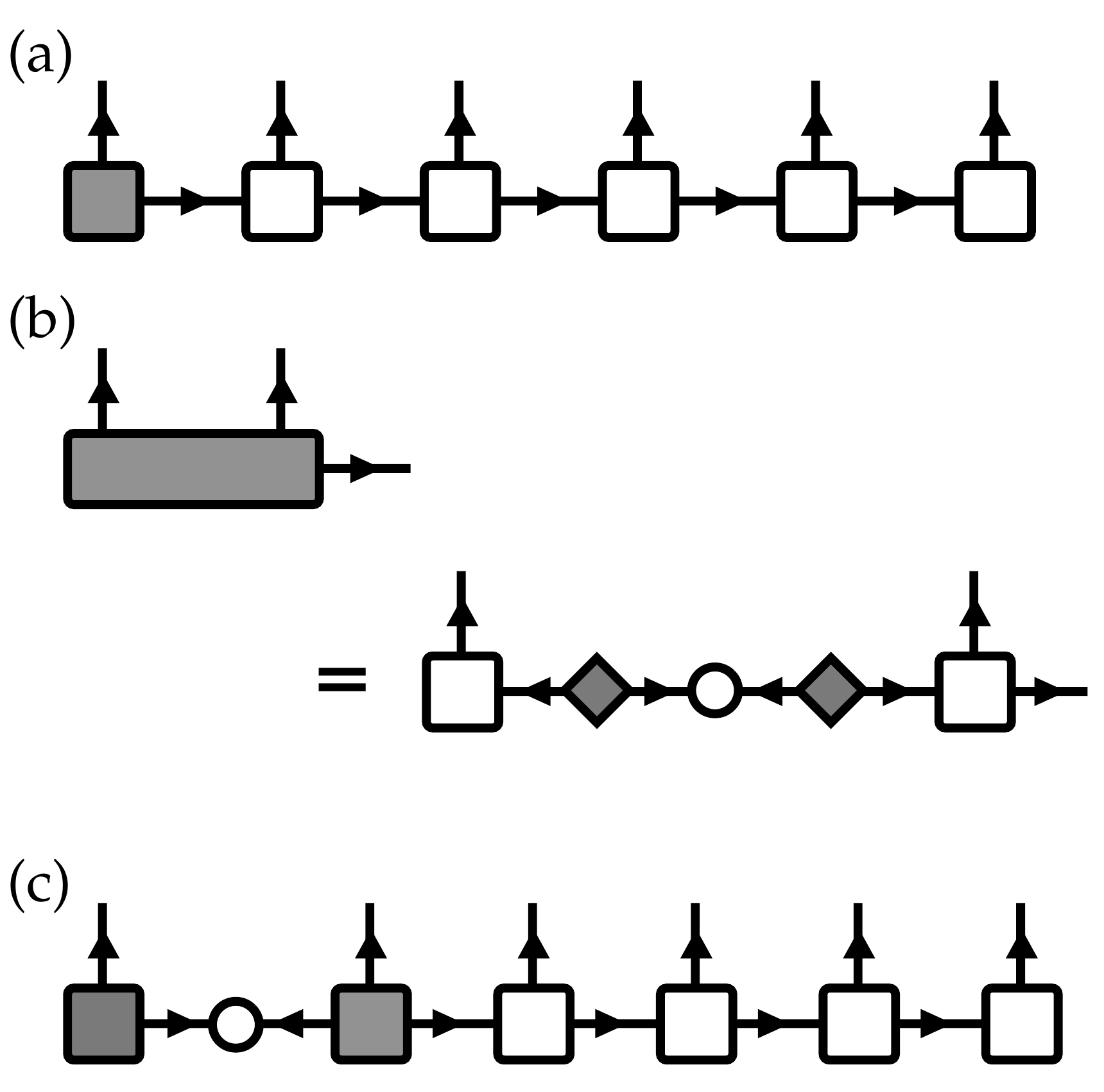}
\caption{Matrix product state in (a) the left-canonical gauge with the first site an OC.
In panel (b) combine the first two site tensors and compute a singular-value decomposition but now with two copies of the 
singular value matrix $\Lambda$ (shaded diamond) and its inverse $V$ (white circle).
Multiplying each singular value matrix $\Lambda$ into the neighboring site tensor
transforms the MPS into (c) a new gauge having two OCs.}
\label{fig:imps_step}
\end{figure}

\begin{figure}[t]
\includegraphics[width=\columnwidth]{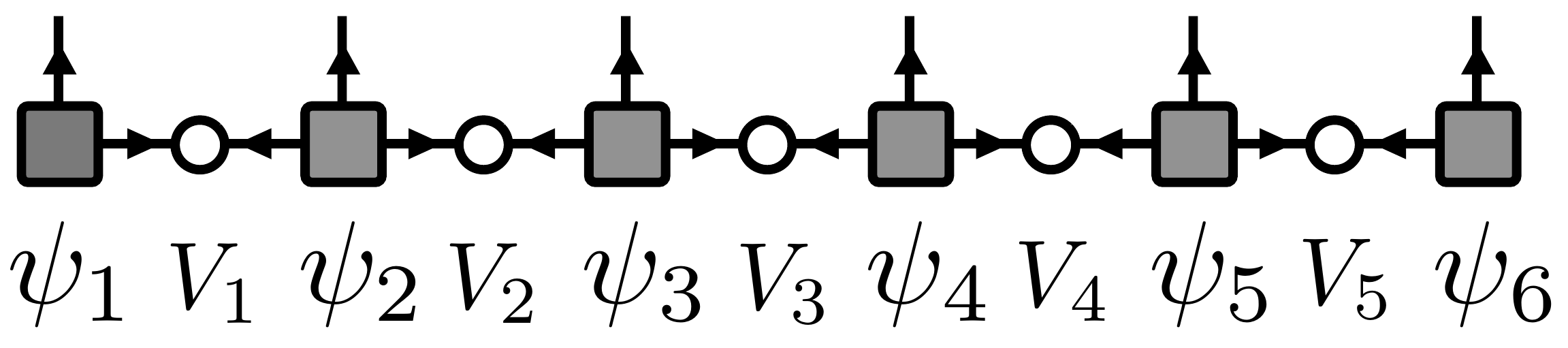}
\caption{Inverse canonical matrix product state gauge. Each site tensor $\psi$ is an OC. 
This gauge is similar to the canonical gauge ($\Gamma$-$\Lambda$ form)\cite{Vidal:2003, Schollwoeck:2011} 
where each bond tensor $\Lambda$ is an OC and is a diagonal matrix containing the Schmidt decomposition weights at that bond.
Here each bond tensor $V=\Lambda^{-1}$ is a diagonal matrix containing the inverse Schmidt weights.
One can directly map between the two gauges via $V_j = \Lambda^{-1}_j$ and $\psi_j = \Lambda_{j-1} \Gamma_j \Lambda_j$.
}
\label{fig:inversecmps}
\end{figure}

Within the parallel DMRG algorithm, when computing an SVD at the shared bond between two nodes, one duplicates the singular value matrix
$\Lambda$ by also inserting the matrix $V\stackrel{\text{def}}{=} \Lambda^{-1}$ on that bond. Multiplying each copy of $\Lambda$ into its neighboring site tensor
creates an MPS gauge with an additional OC. If we repeat the example of the previous paragraph, but now use this modified 
SVD scheme as shown in Fig.~\ref{fig:imps_step}(b), the result is a gauge in which the first two sites are OCs and a matrix $V$ appears on the first bond.
Repeating this procedure at every bond results in the gauge shown in Fig.~\ref{fig:inversecmps}.
This gauge resembles the canonical gauge but has diagonal matrices containg inverse Schmidt coefficients on each bond.
For this reason we refer to it as the inverse canonical gauge.

A key advantage of working in this gauge is that every site tensor is simultaneously an OC. This makes operations such as computing expectation
values of local operators very simple since only site tensors on which an operator acts non-trivially need to be included in the computation (all other site tensors
cancel by construction since they are external to an OC and therefore represent orthonormal states). 
In fact, such local expectation values can be computed in parallel in this gauge.
By contrast, a mixed-canonical MPS such as Fig.~\ref{fig:mixedcmps}(c) must be re-gauged unless the OC is already 
included in the support of the operator to be measured.

Finally, we note that when conserving abelian quantum numbers it is natural to use the same arrow convention described above to denote quantum number flux. 
In this convention, OCs act as
quantum number sources having non-zero flux whereas in a mixed-canonical gauge like that of Fig.~\ref{fig:mixedcmps}(c), for example, 
all other tensors have zero flux since they enact unitary basis transformations.

\section{Conclusion}

We have presented a straightforward modification of the standard DMRG algorithm which allows it to be
parallelized across real-space blocks, providing a nearly ideal speedup. 
The algorithm differs from serial DMRG only at block
boundaries and can readily be combined with other approaches for parallelizing DMRG.
This algorithm should be especially useful for DMRG studies of 2d lattice models, 
quantum chemical systems, and very large or otherwise difficult 1d models. 

We have also presented a set of best practices for parallel DMRG simulations, such as the 
sweeping pattern suggested in Fig.~\ref{fig:sweeping_pattern}, but there remains considerable freedom in implementing 
the algorithm. For example, in our benchmark applications we typically divided the system into real-space blocks of about 8-20 sites, but in principle the 
 blocks could be as small as two sites, offering maximum parallelization. It is interesting to note that fully converging a 
parallel DMRG calculation in this limit would automatically produce an MPS in the inverse canonical gauge. 

Looking ahead, we expect to see real-space parallelism become a standard tool for accelerating
challenging DMRG calculations since it can be implemented in existing codes.
We also hope this work encourages authors of DMRG-related papers to identify parallel aspects of
their methods even more prominently.

\acknowledgments{This paper is dedicated to Ernie Compton, a gifted science teacher and inspiring role model. We thank Thomas Barthel, Bela Bauer, Bryan Clark, Stefan Depenbrock, Adrian Feiguin, Hong-Chen Jiang, Salvatore Manmana, Ian McCulloch, Ulrich Schollw\"ock, Guifr\'e Vidal, and Zhenyue Zhu for helpful discussions. This work is supported by NSF grant DMR-1161348.}

\appendix

\section*{Appendix: Accurate Singular Value Decomposition Algorithm}

When implementing the parallel DMRG algorithm or working with inverse canonical matrix product states it is essential to 
compute singular value decompositions (SVD) to high accuracy. This is because of the presence of the matrix $V$ having 
the inverse singular values along its diagonal. Typical vendor-provided SVD algorithms (such as SGESVD within LAPACK)
may provide poor relative accuracy for the smallest singular values $\lambda_\alpha$, which then translates into very large errors 
upon computing $\lambda^{-1}_\alpha >\!\!> 1$.

To overcome this problem while maintaining efficiency, we have implemented the following SVD algorithm. Though it calls itself recursively,
its asymptotic cost remains $\sim m n^2$ for an $n\times m$ rectangular matrix $M$ (assuming $n<m$ without loss of generality)
since each recursive call only involves a smaller submatrix.

The algorithm proceeds as follows: 

\begin{enumerate}
\item Compute the SVD of \mbox{$M=A \Lambda B$} using a standard algorithm such as SGESVD or through the eigenvalue decomposition of 
\mbox{$\rho \stackrel{\text{def}}{=} M\,M^\dagger = A \Lambda^2 A^\dagger$} (then computing $B$ by orthogonalizing the columns of $A^\dagger M$).

\item Denote the diagonal elements of $\Lambda$ (the singular values) as $\{\lambda_\alpha | \alpha=1\ldots n\}$. For some predetermined 
threshold $\epsilon > 0$, find the smallest integer $p$ such that $\lambda_p/\lambda_1 < \epsilon$. We have found $\epsilon=10^{-4}$ to be a good choice.

\item If no such $p$ exists, the algorithm has
converged. Return the matrices $A$, $\Lambda$, and $B$ from step 1.

\item If the algorithm has not converged, compute \mbox{$X= A^\dagger M B^\dagger$}, 
but only the last $n-p$ rows and columns such that $X$ is an $(n-p)\times (n-p)$ matrix. 
If the SVD of step~1 could be computed exactly, $X$ would be diagonal and contain the last $(n-p)$ singular values.
In practice, $X$ will only be approximately diagonal due to numerical errors.

\item Recursively repeat the algorithm starting again at step 1 but with $M$ replaced by $X$. Denote the resulting SVD matrices $\tilde A$, $\tilde \Lambda$, and $\tilde B$.

\item Update $A$, $B$, and $\Lambda$ as follows:
\begin{align}
A_{i \alpha} & = \sum_{k=p}^{n} A_{i k} \tilde{A}_{k \alpha} & \alpha=p\ldots n \\
B_{\alpha j} & = \sum_{k=p}^{n}  \tilde{B}_{\alpha k} B_{k j} & \alpha=p\ldots n \\
\lambda^\alpha & = \tilde{\lambda}^{\alpha} & \alpha=p\ldots n
\end{align}
where $\lambda^\alpha$ are the diagonal elements of $\Lambda$.

\item Return the updated SVD matrices $A$, $\Lambda$, and $B$.
\end{enumerate}

Because the SVD method used in step~1 of the algorithm is typically accurate for all but the smallest 
singular values (as defined by the threshold $\epsilon$), by calling the method recursively on a submatrix containing 
only these smallest singular values the algorithm finds all the singular values accurately.

\bibliography{pdmrg}

\end{document}